\documentclass[twoside]{dis07}
\usepackage[latin1]{inputenc}
\usepackage[dvips]{graphicx,epsfig,color}
\usepackage{wrapfig,rotating}
\usepackage{amssymb,amsmath,array}
\usepackage{cite}

\newcommand{\simorderr}{\raisebox{-4pt}{$\, \stackrel{\textstyle <}{\sim} \,$}}
\newcommand{\lf}{\left}
\newcommand{\rg}{\right}

\begin{document}
\title{$x$-Evolution of Phenomenological Dipole Cross Sections}
% and the Balitsky-Kovchegov Equation}

%***********************************************************************
% AUTHORS INFORMATION AREA
%***********************************************************************
\author{Dani\"el Boer, Andre Utermann\thanks{Talk \cite{url} presented by Andre Utermann}~ and Erik Wessels
%
% Optional short acknowledgment: remove next line if non-needed
%
% DO NOT MODIFY THE FOLLOWING '\vspace' ARGUMENT
\vspace{.3cm}\\
%
% Addresses and institutions (remove "1- " in case of a single institution)
Department of Physics and Astronomy,\\
Vrije Universiteit Amsterdam, \\
De Boelelaan 1081, 1081 HV Amsterdam, The Netherlands
}
%***********************************************************************
% END OF AUTHORS INFORMATION AREA
%***********************************************************************

\maketitle

\begin{abstract}
  Deep inelastic scattering at small $x$ can be described very
  effectively using saturation inspired dipole models.  We investigate
  whether such models are compatible with the numerical solutions of
  the Balitsky-Kovchegov (BK) equation which is expected to describe
  the nonlinear evolution in $x$ of the dipole cross section. We find
  that the BK equation yields results that are qualitatively different
  from those of phenomenological studies. Geometric scaling is
  recovered only towards asymptotic rapidities. In this limit the
  value of the anomalous dimension $\gamma(r,x)$ at the saturation
  scale approaches approximately $0.44$, in contrast to the value
  $0.63$ commonly used in the models.

\end{abstract}

%\section{Introduction}

At small $x$, deep inelastic scattering (DIS) can be described as the
scattering of a color dipole, which the photon fluctuates into, off
the proton \cite{Mueller:1989st}. The linear BFKL equation, which
describes the dipole-proton interaction in terms of gluon ladders,
predicts an exponential growth of the corresponding cross section as
$\log 1/x$ increases, potentially violating unitarity.  Hence,
interactions between BFKL gluon ladders may become important, which
leads to a nonlinear evolution approximately described by the
Balitsky-Kovchegov (BK) equation \cite{BK}.  As a consequence of the
nonlinearity, the dipole cross section saturates with decreasing $x$,
thereby offering a resolution to the unitarity problem. The inclusive
HERA data at low $x$ ($x \simorderr 0.01$) could be described well by
a dipole cross section of the form $\sigma = \sigma_0 N_{\rm
  GBW}(r,x)$, where the scattering amplitude $N_{\rm GBW}$ is given by
\cite{GBW}
\begin{equation}
N_{\rm GBW}(r,x) = 1-\exp\left[-\tfrac{1}{4} r^2 Q_s^2(x) \right],
\label{NGBW} 
\end{equation}
${r}$ denotes the transverse size of the dipole, $\sigma_0 \simeq
23\;{\rm mb}$ and the $x$-dependence of the saturation scale is given
by $Q_s(x) = 1\,{\rm GeV}\,(x_0/x)^{\lambda/2}$,
where $x_0 \simeq 3 \times 10^{-4}$ and $\lambda \simeq 0.3$. 
The scattering amplitude depends on $x$ and $r$ through the
combination $r^2Q_s^2(x)$ only, which is known as geometric scaling
and leads to the prediction that the structure function $F_2$ is a
function of $Q^2/Q_s^2(x)$ only. This prediction was checked in a
model independent way \cite{Stasto:2000er} and holds widely even
though the GBW model (\ref{NGBW}) is not applicable at large $Q^2$. It
should be mentioned that the leading order BK equation leads to a
faster evolution in $x$ \cite{MuellerTr} ($Q_s^2(x) \sim 1/x^\lambda$
where $\lambda \simeq 0.9$) than the experimental data seem to favor
($\lambda \simeq 0.3$). This discrepancy can be reduced by introducing
a running coupling constant.

Hadron production in $d$-$Au$ collisions can also
be described by saturation inspired dipole models
\cite{KKT,DHJ1,DHJ2}. However, these data seem to require geometric
scaling violation.  The dipole scattering amplitude modified in this
respect is given by~\cite{KKT,DHJ1,DHJ2}
\begin{equation}
N({r},x) =
1-\exp\left[-\tfrac{1}{4}(r^2 Q_s^2(x))^{\gamma(r,x)}\right]\,.
\label{Ngamma}
\end{equation} 
The exponent $\gamma$ is usually referred to as the ``anomalous
dimension'', although the connection of $N$ with the gluon
distribution may not be clear for all cases considered below.
Following partly \cite{IIM,KKT}, in \cite{DHJ1,DHJ2} a few
requirements were used to determine a parameterization of $\gamma$.
Firstly, one assumes that $\gamma(r,x)$ approaches 1 in the limit
$r\to 0$. Therefore the ``DGLAP'' limit $N \sim r^{2}$ is recovered
for all $x$. Secondly, at the saturation scale, $r=1/Q_s$, $\gamma$
should be constant to ensure geometric scaling in this region. This
constant $\gamma_s$ is chosen to be $\simeq 0.628$. The value of
$\gamma_s$ is motivated by a saddle point analysis of a solution of
the BFKL equation with saturation boundary conditions \cite{MuellerTr}
and also shows up in the traveling wave approach \cite{Munier}.
Thirdly, if one writes $\gamma=\gamma_s+\Delta\gamma$, then
$\Delta\gamma$ should decrease as $1/y$ for $y\to\infty$ at fixed $r^2
Q_s^2$. This ensures that geometric scaling is asymptotically
recovered.  Furthermore, the parameters were adjusted in such a way
that geometric scaling holds approximately for finite $y$ in a growing
region between $Q_s(y)$ and roughly $Q_s^2(y)/\Lambda_{\rm QCD}$. Note
that the parameterization in \cite{DHJ2} is intended to describe
$N(r,x)$ in this so-called extended geometric scaling region only.  To
simplify the procedure of the required Fourier transformation of $N$
(\ref{Ngamma}), $\gamma(r,x)$ was replaced in \cite{DHJ1,DHJ2,KKT} by
$\gamma(1/k,x)$ where $k$ is the transverse momentum of the
scattered parton that will fragment into the final state hadron.

We want to check whether these requirements for $\gamma(r,x)$ are
compatible with the nonlinear evolution of the dipole scattering
amplitude $N$. The BK equation for $N$ reads \cite{BK}
\begin{align}
\nonumber
\frac{\partial N\lf(r=|\vec{x}_t-\vec{y}_t|,x\rg)}{\partial y} = & 
\frac{\bar{\alpha}_s}{2\pi} \int d^2 z_t
\frac{(\vec{x}_t-\vec{y}_t)^2}{(\vec{x}_t-\vec{z}_t)^2
(\vec{y}_t-\vec{z}_t)^2} \Big[
N\lf(|\vec{x}_t-\vec{z}_t|,x\rg)+N\lf(|\vec{z}_t-\vec{y}_t|,x\rg) \\[2 mm]
&
-N\lf(|\vec{x}_t-\vec{y}_t|,x\rg)-N\lf(|\vec{x}_t-\vec{z}_t|,x\rg)
N\lf(|\vec{z}_t-\vec{y}_t|,x\rg) \Big].
\label{BK}
\end{align}
Here $\bar{\alpha}_s = \alpha_s N_c/\pi$. We will not consider the
impact parameter dependence of $N$. 

\begin{figure}[htb]
\centering
\includegraphics*[width=69mm]{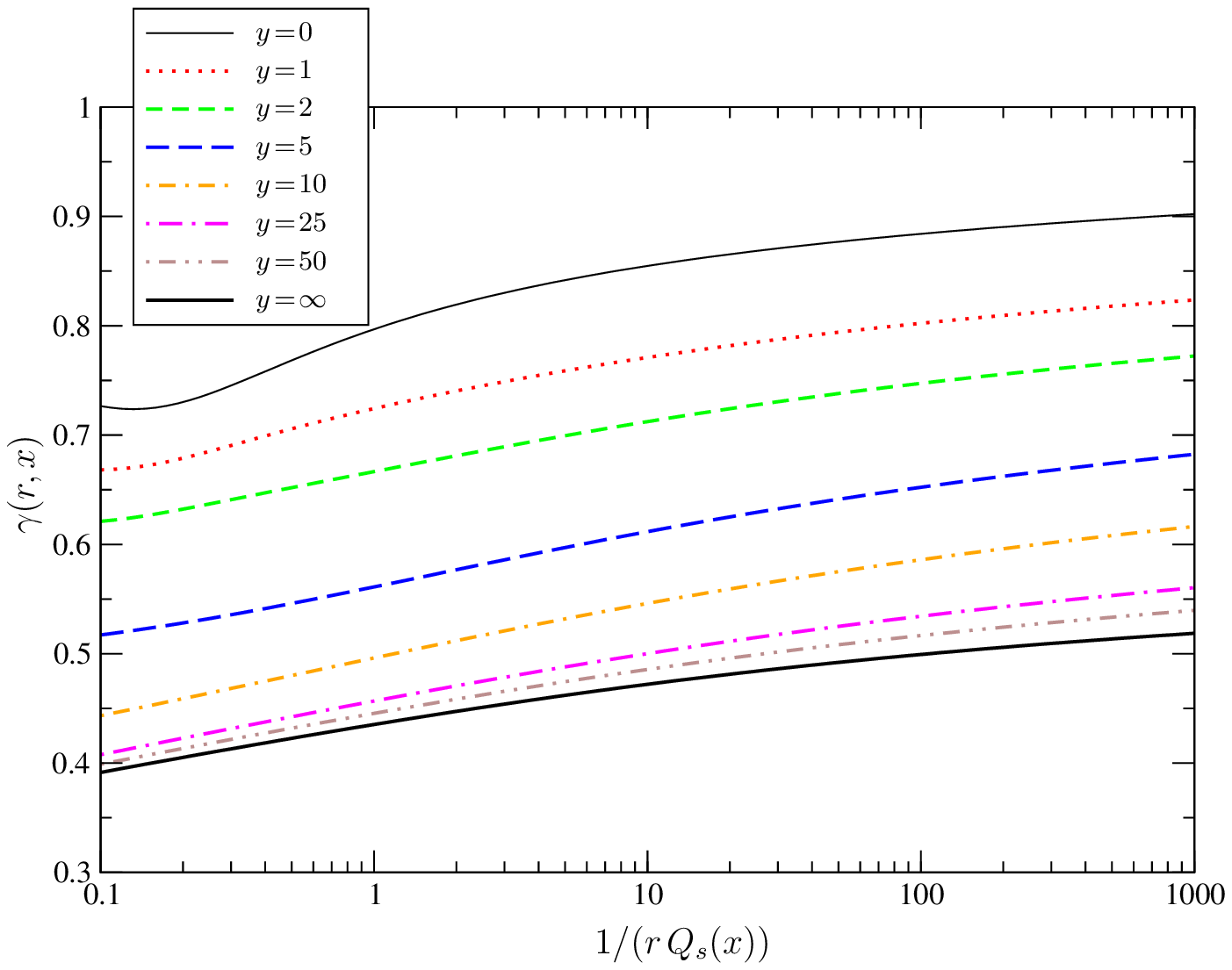}
\includegraphics*[width=69mm]{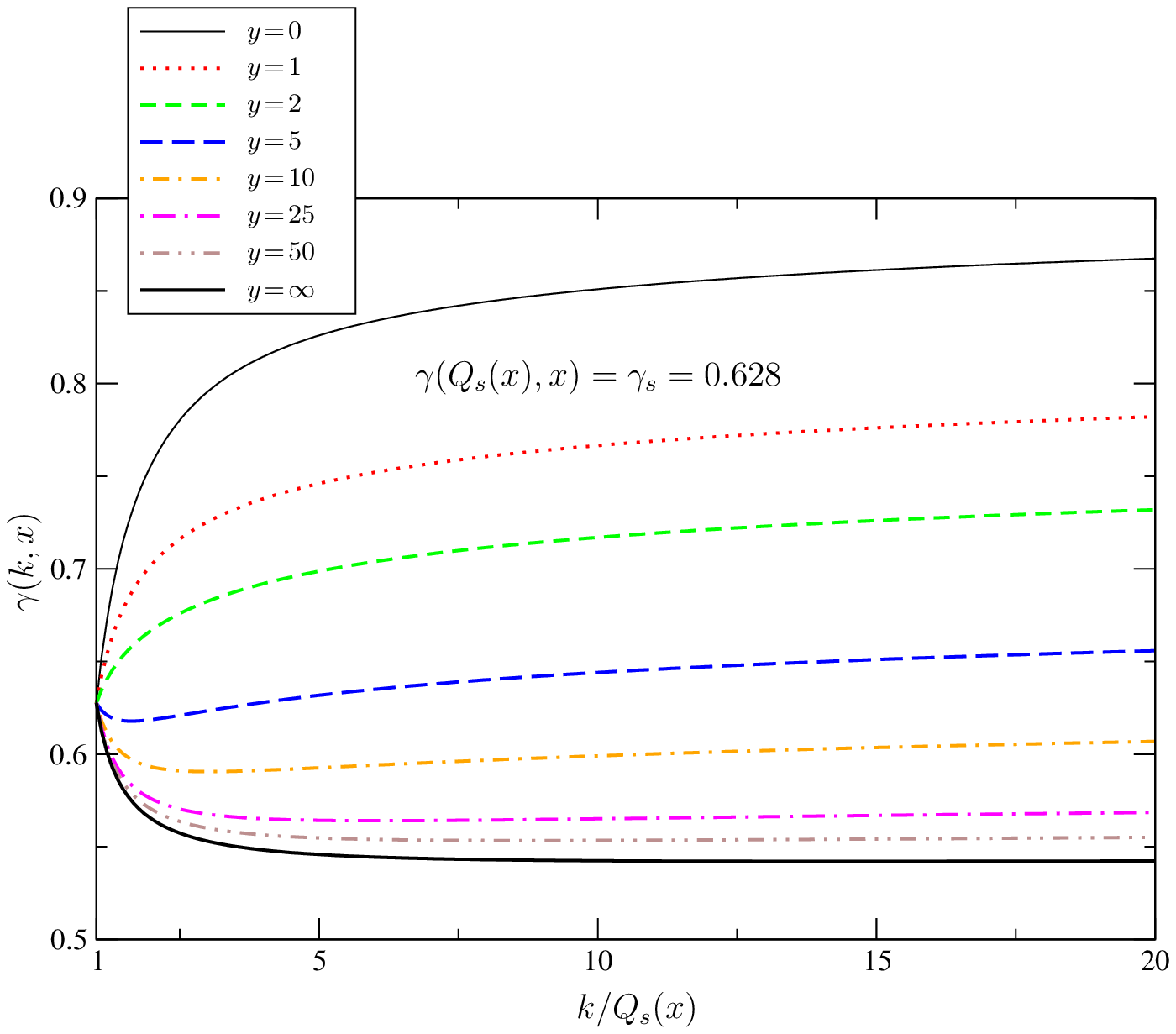}
\caption{\label{gammar1_fig} a) $\gamma(r,x)$ resulting from
  the relations (\ref{Fourier}) and  (\ref{gammar}) as
  a function of $1/(r Q_s(x))$ and $y=\log x_0/x$.
b) $\gamma(k,x)$ as a function of
  $k/Q_s(x)$ for various rapidities $y=\log x_0/x$.}
\end{figure}

%\section{Results}
The BKsolver program \cite{BKSolver} provides a numerical solution of
the amplitude ${\cal N}(k,x)$  in momentum space. In
order to use this solution of the BK equation (\ref{BK}) to constrain
$\gamma(r,x)$, one first has to find $N(r,x)$ by Fourier transforming
to coordinate space:
\begin{equation}
  N(r,x)\equiv r^2\int\frac{d^2
    k_t}{2\pi}\:e^{-i\vec{k}_t\cdot\vec{r}_t}\,\mathcal{N}(k,x)
=r^2\int_0^\infty dk\,k\:{J}_0(k r)\,\mathcal{N}(k,x)\,.\label{Fourier}
\end{equation}
Using the Ansatz (\ref{Ngamma}) one can extract $\gamma(r,x)$ from
the resulting $N(r,x)$,
\begin{equation}
 \gamma(r,x)=\log[\log[(1-N(r,x))^{-4}]]/
\log[r^2\,Q_s^2(x)]\,.
\label{gammar}
\end{equation}
This equation requires as a separate input the value of $Q_s(x)$,
which can be found by equating the right hand sides of
Eqs.~(\ref{Ngamma}) and (\ref{Fourier}) for $r=1/Q_s$. Combining the
resulting values of $Q_s$ with Eq.\ (\ref{gammar}), we obtain a
numerical result for $\gamma(r,x)$, which is shown in
Fig.~\ref{gammar1_fig}a.

The resulting $\gamma(r,x)$ has the following features:
\begin{enumerate}
\item For $r \to 0$, $\gamma(r,x)$ asymptotically approaches 1.
\item At the saturation scale, $\gamma(r,x)$ is not a constant.
\item For decreasing $x$, $\gamma(r,x)$ approaches a limiting curve, 
$\gamma_\infty(rQ_s(x))$, indicated in Fig.~\ref{gammar1_fig} by $y=\infty$. 
Hence, after a longer evolution one indeed recovers geometric scaling.
\end{enumerate}

The fact that for small distances $\gamma$ asymptotically approaches 1
is understandable from the BK equation, since in this limit it reduces
to the BFKL equation.  In the limit of small distances, the solution
to the BFKL equation is dominated by either the saddle point or the
initial condition, both leading to $\gamma\to1$, since here we use the
MV model as the initial condition, see \cite{buw} for details.

It turns out that $\gamma$ is clearly not constant, not even at
$r=1/Q_s$, unlike in \cite{DHJ1,DHJ2,KKT}.  However, asymptotically
geometric scaling is recovered as $\gamma$ approaches $\gamma_\infty$.
Writing
\begin{equation}
 \gamma(r,x)=\gamma_\infty(rQ_s(x))+\Delta\gamma(r,x)\,,
\end{equation}
it turns out that, similar to the parameterizations used in
\cite{DHJ1,DHJ2,KKT}, $\Delta\gamma(r,x)$
decreases as $1/y$ for $y \to \infty$ and fixed $rQ_s(x)$. At the
saturation scale $\gamma$ is given in the small-$x$ limit by
\begin{equation}
\lim_{x\to 0} \gamma(r=1/Q_s(x),x)=\gamma_\infty(1)\approx 0.44\,,
\label{gammars}
\end{equation}
which is significantly below $\gamma_s=0.628$. This is not in
disagreement with theoretical expectations \cite{MuellerTr,Munier}.
Rather it indicates that requiring $\gamma$ in Eq.~(\ref{Ngamma})
to be constant at $Q_s$ does not follow from the BK equation.
 
In \cite{DHJ1,DHJ2,KKT}, $N(r,x)$~(\ref{Ngamma}) was considered with
$\gamma(r,x)$ replaced by $\gamma(1/k,x)$. This approximation scheme
we will discuss next.  The procedure of extracting $\gamma$ becomes
quite different when $\gamma$ depends on $k$, since the dipole cross
section $N$ then depends on both $r$ and $k$, so that it is not
related to $\mathcal{N}(k,x)$ by a straightforward inverse Fourier
transform (\ref{Fourier}) anymore:
\begin{equation}
 \mathcal{N}(k,x) \equiv \int_0^\infty \frac{dr}{r}\:{
   J}_0(kr)\,\lf(1-\exp\lf[-\tfrac{1}{4}(r^2Q^2_s(x))^{\gamma(k,x)}\rg]\rg)\,.
\label{Nk2}
\end{equation}
Instead of by using the inverse Fourier transform, we will extract $\gamma$
by numerically solving Eq.~(\ref{Nk2}), imposing the
following condition. In order to test the Ansatz in \cite{DHJ1,DHJ2},
we will fix $\gamma(k,x)$ in such a way that it equals the constant
$\gamma_s\approx 0.628$ at the saturation scale.  The $x$-dependence
of $Q_s$ is determined by explicitly solving Eq.~(\ref{Nk2}) for
$k=Q_s$ and $\gamma(Q_s,x)=0.628$. Now we can extract $\gamma$ from
relation (\ref{Nk2}) for any given value of $x$ and $k$.
Fig.~\ref{gammar1_fig} shows the results for $\gamma(k,x)$ as a
function of $k/Q_s$ above $Q_s$, for a broad range of rapidities.  For
small rapidities the resulting $\gamma$ looks very similar to the one
in \cite{DHJ2} (cf.\ Fig.\ 4 of Ref.\ \cite{DHJ2}). As one can see,
for larger $y$ the resulting $\gamma$ is not compatible with the
parameterization in \cite{DHJ2} anymore; it first decreases before it
rises towards 1 asymptotically.

For a discussion of additional important issues
like the dependence on initial conditions, the $x$-dependence
of $Q_s$ in our approach and the running coupling case we refer to
\cite{buw}.

\section*{Discussion \& Conclusions}

The numerical solutions of the BK equation do not display exact
geometric scaling, although they approach a solution showing such
scaling at asymptotic $y$. Assuming the solutions to be of the form
(\ref{Ngamma}), where scaling violations are encoded in the
``anomalous dimension'' $\gamma$, therefore leads to the conclusion
that $\gamma(r,x)$ is not a function of $rQ_s(x)$ exclusively. In
particular, it is never simply a constant, not even at the saturation
scale ($r=1/Q_s$). At asymptotically large rapidities, $\gamma$
reaches a limiting function $\gamma_\infty(rQ_s(x))$. This function is
universal for a large range of initial conditions \cite{buw}.  At the
saturation scale, $\gamma_\infty$ equals approximately $0.44$, which is
considerably smaller than the corresponding values in the
phenomenological models \cite{IIM,KKT,DHJ2}. For small values of
$rQ_s$ the limiting function seems to reach $\gamma_s$ \cite{buw}, in
accordance with the traveling wave results of Refs.\ \cite{Munier}.

Performing the replacement of $\gamma(r,x) \to \gamma(1/k,x)$ does
allow one to find a solution for which $\gamma(k=Q_s,x)$ is kept
fixed. The behavior of $\gamma(1/k,x)$ is then for small rapidities
qualitatively similar to the parameterization in \cite{DHJ2}. However,
the usually considered choice $\gamma(k=Q_s,x)=\gamma_s=0.628$ yields
some unwanted features, i.e.\ $\Delta\gamma $ being negative in a
region above the saturation scale and the absence of solutions below
the saturation scale, although the Ansatz was not intended for that
region. Keeping $\gamma(k=Q_s,x)$ fixed at a smaller value, e.g.\ at
$\gamma_\infty(rQ_s=1) \approx 0.44$, seems more suitable \cite{buw},
but it remains to be investigated whether such a choice allows for a
good fit of all relevant DIS, $d$-$Au$ and $p$-$p$ data.

It would be interesting to consider modifications of phenomenological
models for the dipole scattering amplitude that are compatible with
both the BK equation and the data. Given the fact that the BK
evolution does not respect geometric scaling around $Q_s$,
phenomenological parameterizations that reflect this feature would
seem a natural choice. Fortunately, the LHC and a possible future
electron-ion collider will provide data over a larger range of momenta
and rapidities, so that one can expect to test the evolution
properties of the models more accurately.

% ****************************************************************************
% BIBLIOGRAPHY AREA
% ****************************************************************************

\begin{footnotesize}
% IF YOU DO NOT USE BIBTEX, USE THE FOLLOWING SAMPLE SCHEME FOR THE REFERENCES
% ----------------------------------------------------------------------------

% ----------------------------------------------------------------------------

% IF YOU USE BIBTEX,
% - DELETE THE TEXT BETWEEN THE TWO ABOVE DASHED LINES
% - UNCOMMENT THE NEXT TWO LINES AND REPLACE 'Name_Of_Your_BibFile'

%\bibliographystyle{unsrt}
%\bibliography{Name_Of_Your_BibFile}

\begin{thebibliography}{99}
% Please replace the numbers for   contribId   and   sessionId
% in the following URL. You can get this information by going to 
% http://indico.cern.ch/confAuthorIndex.py?confId=9499
% and search for your contribution and click on the title
% Be aware: '&amp;' must be replaced by simple '&' as in example below
\bibitem{url} Slides: \\
\verb$http://indico.cern.ch/contributionDisplay.py?contribId=69&sessionId=15&confId=9499$
%------- replace following references ;-)
\bibitem{Mueller:1989st}
  A.~H.~Mueller,
  %``SMALL x BEHAVIOR AND PARTON SATURATION: A QCD MODEL,''
  Nucl.\ Phys.\ B {\bf 335}, 115 (1990).
  %%CITATION = NUPHA,B335,115;%
\bibitem{BK}
%\bibitem{Balitsky:1995ub}
  I.~Balitsky,
  %``Operator expansion for high-energy scattering,''
  Nucl.\ Phys.\ B {\bf 463}, 99 (1996);
  %%CITATION = HEP-PH 9509348;%%
%\bibitem{Kovchegov:1999yj}
  Y.~V.~Kovchegov,
  %``Small-x F2 structure function of a nucleus including multiple pomeron
  %exchanges,''
  Phys.\ Rev.\ D {\bf 60}, 034008 (1999).
  %%CITATION = HEP-PH 9901281;%%

\bibitem{GBW}
  K.~Golec-Biernat and M.~W\"usthoff,
  Phys.\ Rev.\ D {\bf 59}, 014017 (1999).
  %%CITATION = HEP-PH 9807513;%%

\bibitem{Stasto:2000er}
  A.~M.~Stasto, K.~Golec-Biernat and J.~Kwiecinski,
  Phys.\ Rev.\ Lett.\  {\bf 86}, 596 (2001).
  %%CITATION = HEP-PH 0007192;%
%
\bibitem{MuellerTr}
  A.~H.~Mueller and D.~N.~Triantafyllopoulos,
  %``The energy dependence of the saturation momentum,''
  Nucl.\ Phys.\ B {\bf 640}, 331 (2002).
  %%CITATION = HEP-PH 0205167;%%

\bibitem{KKT}
  D.~Kharzeev, Y.V.~Kovchegov and K.~Tuchin,
  Phys.\ Lett.\ B {\bf 599}, 23 (2004).
  %%CITATION = HEP-PH 0405045;%%

\bibitem{DHJ1}
  A.~Dumitru, A.~Hayashigaki and J.~Jalilian-Marian,
  %``The color glass condensate and hadron production in the forward region,''
  Nucl.\ Phys.\ A {\bf 765}, 464 (2006).
  %%CITATION = HEP-PH 0506308;%%
\bibitem{DHJ2}
  A.~Dumitru, A.~Hayashigaki and J.~Jalilian-Marian,
  %``Geometric scaling violations in the central rapidity region of d + Au
  %collisions at RHIC,''
  Nucl.\ Phys.\ A {\bf 770}, 57 (2006).
  %%CITATION = HEP-PH 0512129;%%

%\bibitem{IIM2}
%  E.~Iancu, K.~Itakura and L.~McLerran,
%``Geometric scaling above the saturation scale,''
%  Nucl.\ Phys.\ A {\bf 708}, 327 (2002).
%%CITATION = HEP-PH 0203137;%%s

\bibitem{IIM}
  E.~Iancu, K.~Itakura and S.~Munier,
  Phys.\ Lett.\ B {\bf 590}, 199 (2004).
  %%CITATION = HEP-PH 0310338;%%

\bibitem{Munier}
  S.~Munier and R.~Peschanski,
  %``Geometric scaling as traveling waves,''
  Phys.\ Rev.\ Lett.\  {\bf 91}, 232001 (2003);
  %%CITATION = HEP-PH 0309177;%%
%\bibitem{Munier:2003sj}
%  S.~Munier and R.~Peschanski,
  %``Traveling wave fronts and the transition to saturation,''
  Phys.\ Rev.\ D {\bf 69}, 034008 (2004).
  %%CITATION = HEP-PH 0310357;%%

\bibitem{BKSolver}
  R.~Enberg, ``BKsolver: numerical solution of the Balitsky-Kovchegov
  nonlinear integro-differential equation'', available at URL:
  http://www.isv.uu.se/$\sim$enberg/BK/. 

\bibitem{buw}
  D.~Boer, A.~Utermann and E.~Wessels,
  %``Compatibility of phenomenological dipole cross sections with the
  %Balitsky-Kovchegov equation,''
  Phys.\ Rev.\  D {\bf 75}, 094022 (2007).
%  [arXiv:hep-ph/0701219].
  %%CITATION = PHRVA,D75,094022;%%

\end{thebibliography}
% example of Name_Of_Your_BibFile.bib
% @Article{Turcato:2006ch,
%      author    = "Turcato, M.",
%  collaboration = "ZEUS and H1",
%      title     = "Lepton flavour violation and charmonium physics at HERA",
%      journal   = "Nucl. Phys. Proc. Suppl.",
%      volume    = "162",
%      year      = "2006", 
%      pages     = "283-287",
%      SLACcitation  = "%%CITATION = NUPHZ,162,283;%%"
% }
% 
% @Unpublished{Gogitidze:2007du,
%      author    = "Gogitidze, N.",
%  collaboration = "H1", 
%      title     = "Prompt photons and particle momentum distributions at
%                   HERA", 
%      year      = "2007",
%      note    = "hep-ex/0701033",
%      SLACcitation  = "%%CITATION = HEP-EX 0701033;%%"
% }

\end{footnotesize}

% ****************************************************************************
% END OF BIBLIOGRAPHY AREA
% ****************************************************************************

\end{document}